\documentstyle[epsfig,11pt]{article}
\title{\bf Elliptic flow from colour strings}
\author{M.A.Braun$ ^{a,b}$, C.Pajares$^a$\\
 $^a$ University of Santiago de Compostela, Spain,\\
$^b$ S.Petersburg State University, Russia
}

\begin{document}
\maketitle
\input epsf
\def\beq{\begin{equation}}
\def\eeq{\end{equation}}

{\bf Abstract}
It is  shown that the elliptic flow can be successfully described in the colour string picture with fusion and percolation provided  anistropy of 
particle emission from the fused string is taken into account. Two possible 
sources of this anisotropy are considered, propagation of the string in the transverse plane and quenching of produced particles in the strong colour field of the string. Calculations show that the second source gives an overwhelming contribution to the flow at accessible energies.


\section{Introduction}
The observed elliptic flow in  heavy-ion collisions can be conveniently 
understood if the distribution of the observed particles depends not only
on the physical conditions realized locally at their production point but also on the global geometry of the event. In a relativistic local theory this non-local information can only emerge as a collective effect, requiring interaction between the relevant degrees of freedom localized at different points in the collision region. In this sense anisotropic flow is a particularly unambiguous and strong manifestation of collective dynamics in heavy-ion collisions. The large elliptic flow coefficient $v_2$ can be qualitatively understood as follows. In a high-energy collision spectator nucleons are fast enough to move away leaving behind at mid-rapidity an almond shaped azimuthally asymmetric overlap region filled
with the QCD matter. This spatial asymmetry implies unequal pressure gradients in the transverse plane, with a larger density gradient perpendicular to the 
reaction plane (in-plane). As a consequence of the subsequent multiple interactions between the degrees of freedom involved, the spatial asymmetry leads to anisotropy in the momentum space \cite{borghini}. The final particle transverse momenta are more likely to lie in the in-plane than in the perpendicular direction, which leads to $v_2>0$ as predicted in \cite{ollitraut}.

This general idea has been realized in various mechanisms for the source of elliptic flow. A convenient and successful way to describe the flow anisotropy is achieved in the hydrodynamical approach, taking into account the unsymmetric
shape of the nuclei overlap in the transverse plane at values of the impact parameter ${\bf b}$ different from zero. This description of course assumes collective effects to be responsible for the flow. The flow can also be explained in 
 different terms as following from the quenching of initially produced particles in the nuclear matter. Since the path inside the nuclear overlap of particles emitted along  ${\bf b}$ and orthogonal to it is different, the observed dependence on the azimuthal angle $\phi$ of emission is natural.
Finally in \cite{kaidalov} the authors pointed out that the origin of the flow may be already contained in the asymmetry of the initial emission,
before any collective effects have taken place. They used the Regge picture of particle production and showed that the flow  may follow from the dependence
of the effective emission vertex on the angle between the directions of the
emitted particle momentum ${\bf p}$ and pomeron propagation ${\bf e}$. In the end the strength of the flow, characterized by the standard coefficient $v_2$ was related to the model dependent coefficient $\epsilon$ in the dependence on
$({\bf pe})^2$.

In this note we want to draw attention that the elliptic flow can also be 
naturally explained in the colour string approach,
duly  generalized to include string fusion and percolation \cite{armesto,
braun1,braun2,braun3}. This approach
has proved to be
quite successful in the description of particle production and correlations in the soft part of spectra \cite{braun4,DDD,cunqueiro}. In the original formulation in which strings were assumed to be just points in the transverse space there was no place for elliptic flow whatever the string distribution in the nuclei overlap.
However such a picture is of course too simplified. It neglects at least two 
circumstances. First, strings can propagate in the
transverse space and so be characterized by a vector in this space rather
than a point. In the Regge language this corresponds to assuming that the pomeron slope $\alpha'$ is substantially different from zero, in contrast to the original assumption that the slope is quite small. In the 3-dimensional space
the string becomes not orthogonal to the transverse plane and one may expect an anisotropic emission of particles in this plane and a non-zero elliptic flow.
This mechanism is quite similar to the one considered in
 \cite{kaidalov} with the anisotropy created already in the initial emission.
Of course the actual
form of this anisotropy cannot be established on purely theoretical grounds but has to be taken in a semiphenomenological manner.
We study possible forms of this anisotropy and compare it with the experimental situation. Our conclusion is that with any choice of anisotropy in the
emission of the string the final elliptic flow is far too small as compared
to the experimental data. So our conclusion is that taking into account
only the flow from the initial emission, one cannot explain the data.
Inclusion of collective effects in the form of consecutive 
interactions seems to be anavoidable.

This motivated us to study an alternative mechanism of azimuthal anisotropy,
using the second circumsance characteristic for a realistic string picture,
namely that the string is not point-like in the transverse plane but occupies a certain non-zero area in the transverse space and that a fused string occupies a greater area than the simple string.
Then partons emitted at some point inside the string have to pass a certain length before they appear outside and are observed. It is natural to assume that as they pass through the strong colour field inside the string they emit gluons
 and their energy diminishes. As a result, the particle observed with transverse momentum $p$ outside the string has to be born with a greater momentum inside the string, whose value depends on the path lengh travelled inside the string and so different for different direction of emission. In its spirit, this mechanism is similar to  that of particle quenching in the nuclear matter giving the
string explanation and characteristics of this quenching.
Our calculations confirm that a sizable elliptic flow follows from this mechanism. Its centrality and transverse momentum dependence agrees with the behaviour of the RHIC data \cite{adler,adare,alver,voloshin}. These results also confirm the ones obtained in a similar framework using different simplified methods 
\cite{bautista1,bautista2}

\section{A string stretched between points $ \beta_1$ and $\beta_2$}
We  consider a model in which strings can be formed between different points in the transverse plane,
say, $ \beta_1$ in the projectile nucleus and $\beta_2$ in the target one. 
Here $\beta_{1,2}$ are in fact two-dimensional vectors, which  we do not specifically mark in the following. The probability to form such a string will
be given by the distribution in $\beta= \beta_1-\beta_2$ of the soft pomeron
\beq
P(y,\beta)=\frac{1}{4\pi\alpha'y}e^{\epsilon y-\beta^2/(4\alpha'y)}.
\eeq
Obviously the pomeron extends to larger distances as $y$ grows, the average distance being
$<\beta>=\sqrt{4\alpha' y}$. This will be also the average length of the string in the transverse plane.

\begin{figure}
\epsfig{file=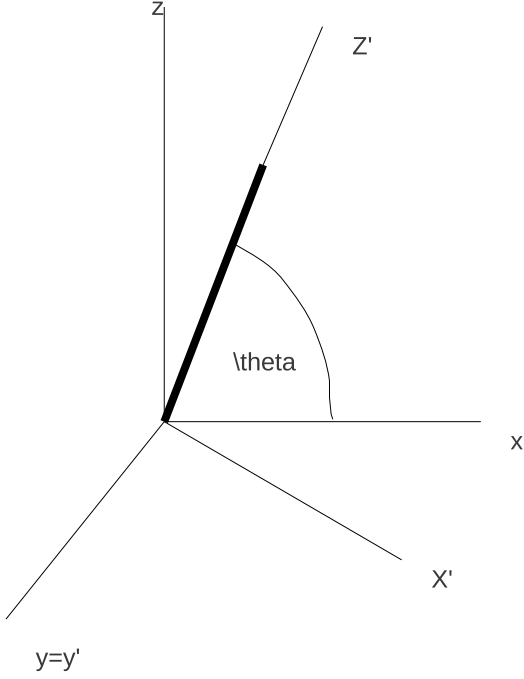,width=10 cm}
\caption{The string in the 3-dimensional space. The shown angle is
 $\theta=\pi/2-\theta'$ }
\label{fig1}
\end{figure}

We shall be interested in the distribution of particles emitted from such a string. To see it we have to recall that the string has also some dimension along the $z$-axis. We assume that the string is located in the $xz$-plane and its direction defines the $z'$ axis in the primed coordinate system $(x',y',z')$. 
The $y'$ axis is assumed to coincide with the $y$ axis in the original system 
and the $x'$ axis is taken to be orthogonal to $y'$ and $z'$ axes
 (see Fig.\ref{fig1}).

Our initial assumption, borrowed from the Schwinger mechanism of particle emission in the external field, is that particles
are emitted isotropically only in the plane orthogonal to the string direction. So in the primed system we find
that the probabilty to emit a particle with momentum $p'$  is
\beq
\mu'(y,p')=\delta(p'_z)\frac{\xi}{\pi}e^{-\xi ({p'}_x^2+{p'}_y^2)}.
\eeq
To find the distribution in the original system we have to express $p'$ via the momentum $p$ in
the original system, that is to make a rotation by angle $\theta'$
between the two systems.
We find the distribution of emitted particles in $p_\perp$ in the original  system as
\beq
\mu(y,p_\perp)=\frac{\xi}{\pi\cos\theta'}e^{-\xi (p_y^2+p_x^2/\cos^2\theta')}.
\eeq

Angle $\theta'$ is given by
\beq
\theta'=\arctan\frac{\beta}{\Delta z}.
\eeq
As we have seen, on the average $\beta\sim\sqrt{\alpha' y}$ and grows with energy of the string $W=e^{y/2}$.
The length in $z$ can be estimated as $1/W= e^{-y/2}$. So angle $\theta'$ is 
quite close to $\pi/2$. Putting
$
\theta'=\frac{\pi}{2}-\theta
$ we find that
$\theta$ is small and diminishes with energy:
\beq
\theta\simeq \frac{\Delta z}{\beta}\sim\frac{e^{-y/2}}{\sqrt{\alpha' y}}.
\eeq
This means that unless $\alpha'=0$ the string only radiates in the direction of $y$,
so that
\beq
\mu(y,p_\perp)=\sqrt{\frac{\xi}{\pi}}\delta(p_x)e^{-\xi p_y^2}=
\sqrt{\frac{\xi}{\pi}}\delta(p_x)
e^{-\xi (p_\perp^2-\frac{(p_\perp \beta)^2}{\beta^2})}.
\eeq
On the other hand, if $\alpha'=0$ then $\theta'=0$, $\theta=\pi/2$ and we return to the old case of string characterized by a single
impact parameter with the emission probability
\beq
\mu(y,p_\perp)_{\alpha'=0}=\frac{\xi}{\pi}e^{-\xi p_\perp^2}.
\eeq
Obviously with $\alpha'\neq 0$ emission is  anisotropic and we expect a non-zero elliptic flow effect.

One has to understand that this ideal case may have little to do with the realistic behaviour of the string,
which does not exactly conform to the Schwinger mechanism,
since its dimensions are finite. To come somehow closer to reality we may soften our assumption that the string
does not emit particles along its axis at all. Instead we may assume that emission along the axis is simply different from emission in the transverse plane,
so that the three-dimensional distribution in the primed system is
\beq
\mu(y,p')=\sqrt{\eta}\Big(\frac{\xi}{\pi}\Big)^{3/2}e^{-\xi(\eta  {p'}_x^2+{p'}_y^2+{p'}_z^2)},
\eeq
where  $\eta$ shows the difference in emission along the string and orthogonal to it.
If $\eta>1$ the emission along the string is damped. 
In the limit $\eta\to\infty$ we return to the previous case of an ideal string. If $\eta<1$ then, on the contrary, emission along the string direction is enhanced.
The same distribution in the original system is easily obtained expressing the primed momenta via 
the ones in the original system.
The final distribution in the transverse momentum $p_\perp$ is obtained after integration over $p_z$ and we find
\beq
\mu(y,p_\perp)=\int dp_z\mu(y,p)=\frac{\xi\sqrt{\gamma}}{\pi}
\exp\Big(-\xi (p_y^2+\gamma p_x^2)\Big),
\label{mu1}
\eeq
where the final excentricity parameter $\gamma$ is 
\beq
\gamma=\frac{\xi}{\cos^2\theta+\xi\sin^2\theta}
\label{gamma}
\eeq
and we used $\theta'=\pi/2-\theta$

In the high-energy limit when $\theta\to  0$ we now find $\gamma=\eta$ and
\beq
\mu(y,p_\perp)=\frac{\xi}{\pi}\sqrt{\eta}
\exp\Big[-\xi\Big(p_y^2+\eta p_x^2)\Big)\Big].
\eeq
If $\alpha'=0$ we have $\theta=\pi/2$, $\gamma=1$ and 
\beq
\mu(y,p_\perp)=\frac{\xi}{\pi}
\exp\Big[-\xi\Big(p_y^2+p_x^2\Big)\Big].
\eeq	  
Obviously with  $\gamma$ different from unity the anisotropy is  fully determined by its value.

\section{Elliptic flow from independent strings}
Once emission from a string is anisotropic, it is obvious that we shall find a non-zero
elliptic flow in AB scattering. The derivation closely follows \cite{kaidalov}. Let the
probability to find a string attached to point $\beta_{1}$ in the projectile
 nucleus A to be
\beq 
P_{A}(\beta_{1}=\int d^2b_1T_A(b_1)t(\beta_1-b_1),
\eeq
where $T_A$ is the profile function, and a similar
probability $P_B(\beta_2)$ to find a string attached to point $\beta_2$ in the target nucleus
\beq 
P_B(\beta_2)=\int d^2b_1T_B(b-b_2)t(\beta_2-b_2),
\eeq
where $b$ is the
impact parameter of the collision.
The probability to find a string stretched between the points $\beta_1$ and $\beta_2$
will be determined by the factor $p(\beta_1-\beta_2)$, where 
\beq
p(\beta)=\frac{1}{4\pi\alpha' y}\exp\Big(-\frac{\beta^2}{4\alpha' y}\Big).
\eeq
So the final distribution of emiitted particles will be
\beq
M(b,p)=\int d^2\beta_1d^2\beta_2 P_A(\beta_1)P_B(\beta_2)p(\beta_1-\beta_2)
\mu(y,\beta,p_\perp),
\label{muab}
\eeq
where $\mu$ is given by (\ref{mu1}) with the $x$-axis along the direction of the string, that is, 
along $\beta=\beta_1-\beta_2$

One can transform (\ref{muab}) to the form introduced in \cite{kaidalov}:
\beq
M(b,p)=\int d^2\beta T_{int}(b-\beta)\Gamma(\beta,y,p_\perp),
\eeq
where obviously
\beq
T_{int}(b)=\int d^2\beta_1P_A(\beta_1)P_B(\beta-\beta_1)
\eeq
and
\beq
\Gamma(\beta,y,p_{\perp})=p(\beta)\mu(\beta,y,p_\perp),
\eeq
with
\beq
\mu(\beta,y,p_\perp)=\frac{\xi\sqrt{\gamma}}{\pi}
\exp\Big[-\xi\Big(p^2-\frac{(p\beta)^2}{\beta^2}+
\gamma\frac{(p\beta)^2}{\beta^2}
\Big)\Big].
\label{mubeta}
\eeq

Taking the final $x$-axis along the direction of $p_\perp$ we get the expression 
which can be compared with \cite{kaidalov}
\beq
\Gamma(\beta,p_\perp)=\frac{\xi\sqrt{\gamma}}{4\pi^2\alpha' y}
\exp\Big[-\frac{\beta^2}{4\alpha' y}-\xi p^2\Big(\sin^2\chi+
\gamma\cos^2\chi\Big)\Big],
\label{mubeta1}
\eeq
where $\chi$ is the azimuthal angle between $\beta$ and $p_{\perp}$.

As compared to \cite{kaidalov} the role of $p^2$ and $\beta^2$
are reversed in our approach. In our case the distribution $\Gamma$ is essentially
\beq
\Gamma\sim e^{-\beta^2/\beta_0^2-\xi p^2(\sin^2\chi+\gamma\cos^2\chi)},
\eeq
where $\beta_0^2=4\alpha' y$ and 
 excentricity $\gamma$ is given by Eq. (\ref{gamma})
(it depends (weakly) on the string dimension $\beta$).
In  \cite{kaidalov} one has
\beq
\Gamma\sim e^{-\xi p^2-(\beta^2/\beta_0²)(\sin^2\chi+\gamma\cos^2\chi)},
\eeq
where $\beta_0$ and $\gamma$ are expressed via parameters of the emission vertex
in the Regge approach
\beq
\beta_0=2r_q,\ \ \gamma=\frac{r_q^2}{r_q^2+\epsilon r_0^4p^2},
\eeq
with $r_q\sim r_0\sim 1$ fm. In \cite{kaidalov} $\gamma$ depends 
on $p^2$  (not weakly).

However it turns out  that after integration over $\phi$ both forms are essentially equivalent. The main difference in the results comes from the choice of parameters. In particular with the standard choice of the pomeron slope
$\alpha'=0.2$ GeV$^{-2}$ our $\beta_0$ turns out to be quite small  at
accessible rapidities as compared
to the value assumed in \cite{kaidalov} and
 becomes of the
same magnitude only at $y\sim 100$

\section{Strings homogeneosly distributed}
\subsection{Analytic expressions}
To obtain concrete results we have to specify the distributions
of strings inside the nuclei. For simplicity we assume  strings
to be homogeneously distributed in the  transverse plane inside the
nucleus:
\beq
P_A(\beta_1)= N_s\frac{\theta (R_A-\beta_1)}{\pi R_A^2},
\eeq
where $N_s$ is the number of strings and $R_A$ is the nucleus radius. We also consider a simple case
of a collison of two identical nuclei, so that,  up to coefficient, 
$T_{int}$
is just the overlap area of the nuclei at
distance $\beta$ between their centers:
\beq
T_{int}(\beta)=\frac{N_s^2}{\pi^2 R_A^2}(2\zeta-\sin(2\zeta),\ \ 
\zeta= \arccos\frac{\beta}{2R_A}.
\label{tint}
\eeq

For the distribution in  angle $\phi$ between $p$ and $b$ the
coefficiens in front of the inclusive cross-section are unimportant. So for
our purpose we find
\beq
M(b,p)=\int_0^{\beta_m}\int_0^{2\pi}d\chi (2\zeta-\sin(2\zeta)
e^{-\beta^2/\beta_0^2}
e^{-\xi p^2(\sin^2\chi+\gamma\cos^2\chi)},
\eeq
where 
\beq
\beta_m=2R_A+b,\ \ \beta_0=4\alpha'y,\ \ 
\zeta= \arccos\frac{|b-\beta|}{2R_A},
\eeq
$\gamma$ is given by Eq. (\ref{gamma}).
and $\xi$ is the inverse string tension which determines the distribution in
momenta.
The dependence of $M(b,p)$ on $\phi$ comes from function $\zeta$.
If we put 
$
\chi=\chi'+\phi
$
then
\beq
\zeta=\arccos\frac{\sqrt{b^2+\beta^2-2b\beta\cos\chi'}}{2R_A}
\eeq
and we get
\beq
M(b,p)=\int_0^{\beta_m}\int_{-\pi}^{\pi}d\chi' \Big(2\zeta-\sin(2\zeta)\Big)
e^{-\beta^2/\beta_0^2}
e^{-ap^2(\sin^2(\chi'+\phi)+\gamma\cos^2(\chi'+\phi))},
\eeq
where integration over $\chi'$ should be restricted by the condition
\beq
|\chi'|<\arccos\frac{b^2+\beta^2-4R_A^2}{2b\beta}
\eeq
unless the argument of the arccos is less than $-1$.

\subsection{Numerical results}
We take the standard value for $\alpha'=0.2$ GeV$^{-2}$.
For the string tension $\xi$ we choose $\xi=0.25$  GeV$^{-2}$.
The only left parameter is excentricity $\gamma$.
From the start one finds that with values of $\gamma$ greater than unity
we obtain negative values for the elliptic flow parameter $v_2$ and
with values of $\gamma$ less than unity we get positive values of $v_2$.
Should the elliptic flow come only from the discussed
effect of string extension in the transverse plane, 
the experimental data would exclude $\gamma>1$ and thus the naive picture in 
which the string only emits particles in the plane transverse to its direction, as in the Schwinger picture. Rather we have to admit the opposite: the string emits particles predominantly along its direction, which leads to $\gamma<1$.
This does not look too exotic if the length of the string in the $z$-direction is much smaller than in its transverse direction.

\begin{figure}
\epsfig{file=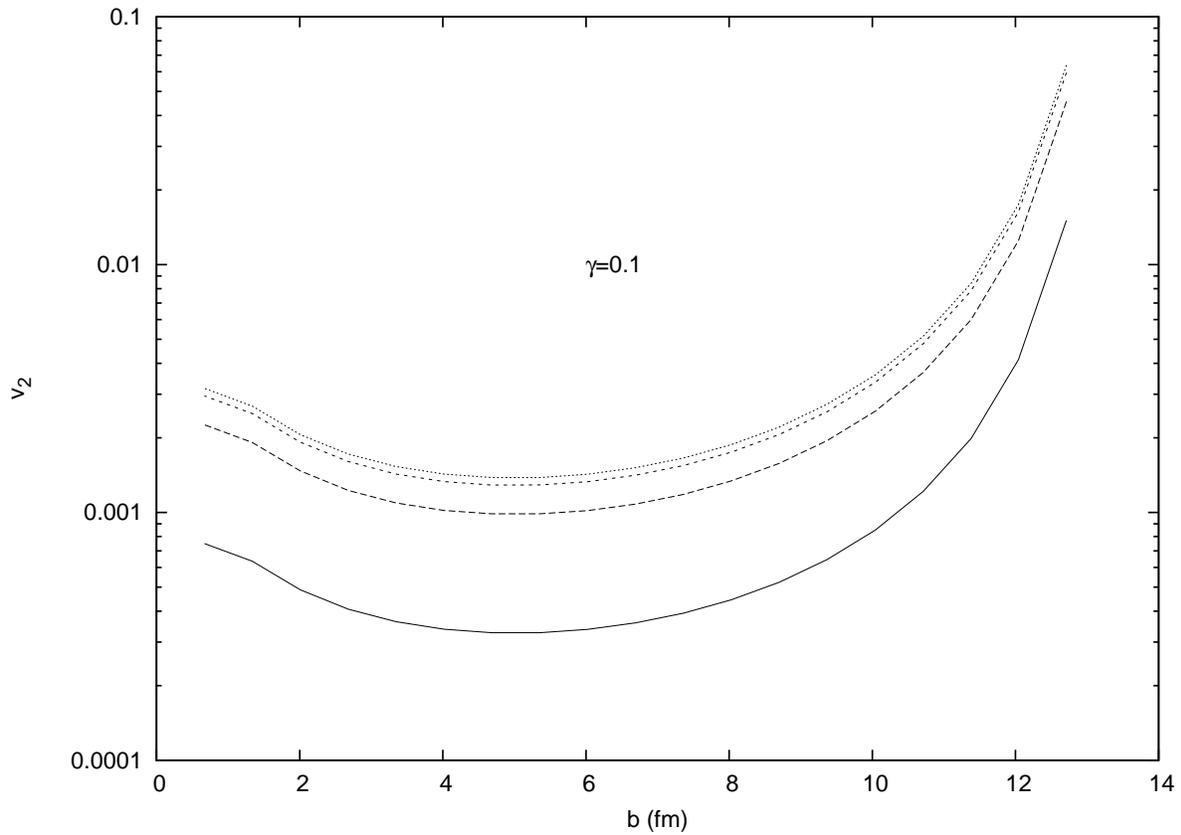,width=16 cm}
\caption{Elliptic flow coefficient $v_2$ with $\gamma=0.1$ as a function of $b$
for Au-Au collisions at $y=10$. The curves from bottom to top correspond to
$p=0.5,\, 1.\, 1.5$ and 2 GeV }
\label{fig2}
\end{figure}

\begin{figure}
\epsfig{file=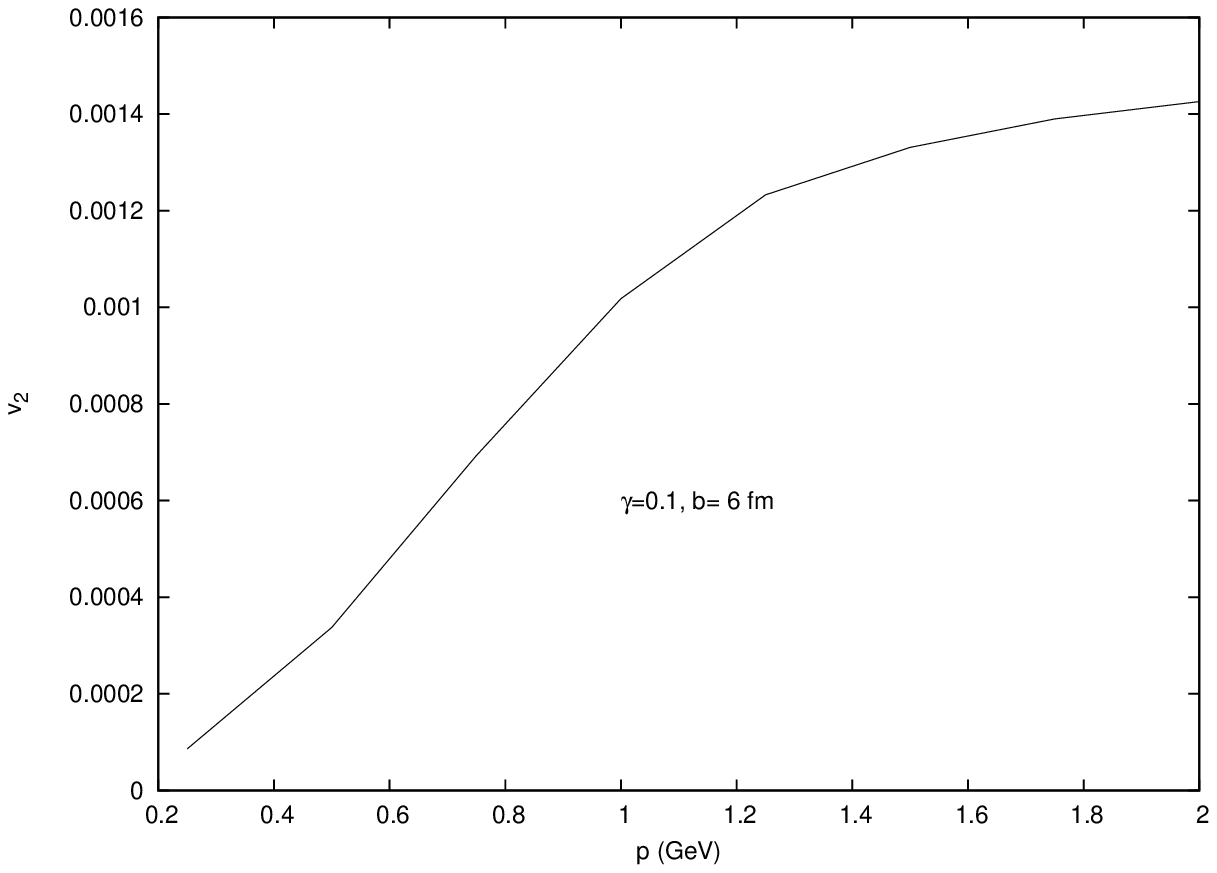,width=16 cm}
\caption{Elliptic flow coefficient $v_2$ with $\gamma=0.1$ as a function of $p$
at $b=6$ fm
for Au-Au collisions at $y=10$.}
\label{fig3}
\end{figure}

However both with $\gamma<1$ and $\gamma>1$ the magnitude of the elliptic 
flow turns out to be quite small at accessible rapidities. We studied Au-Au 
collisiona at rapidity $y=10$. In Fig. \ref{fig2} and \ref{fig3}
 we show our results with $\gamma=0.1$ which lead to positive values for $v_2$.
 In Fig. \ref{fig2} we 
show $v_2$ as a function of impact parameter $b$ at different values of $p$.
 In Fig.\ref{fig3} we illustrate the $p$ dependence of the elliptic flow  
at $b=6$. As we see the qualitative behaviour of both $b$- and $p$- dependence
of the elliptic flow agrees with the experimental findings. However values of
$v_2$ at not too peripheral collisions are ten to twenty times lower than the
 data. Only in the limiting case of a very small overlap these values
become comparable to the observed ones.

\begin{figure}
\epsfig{file=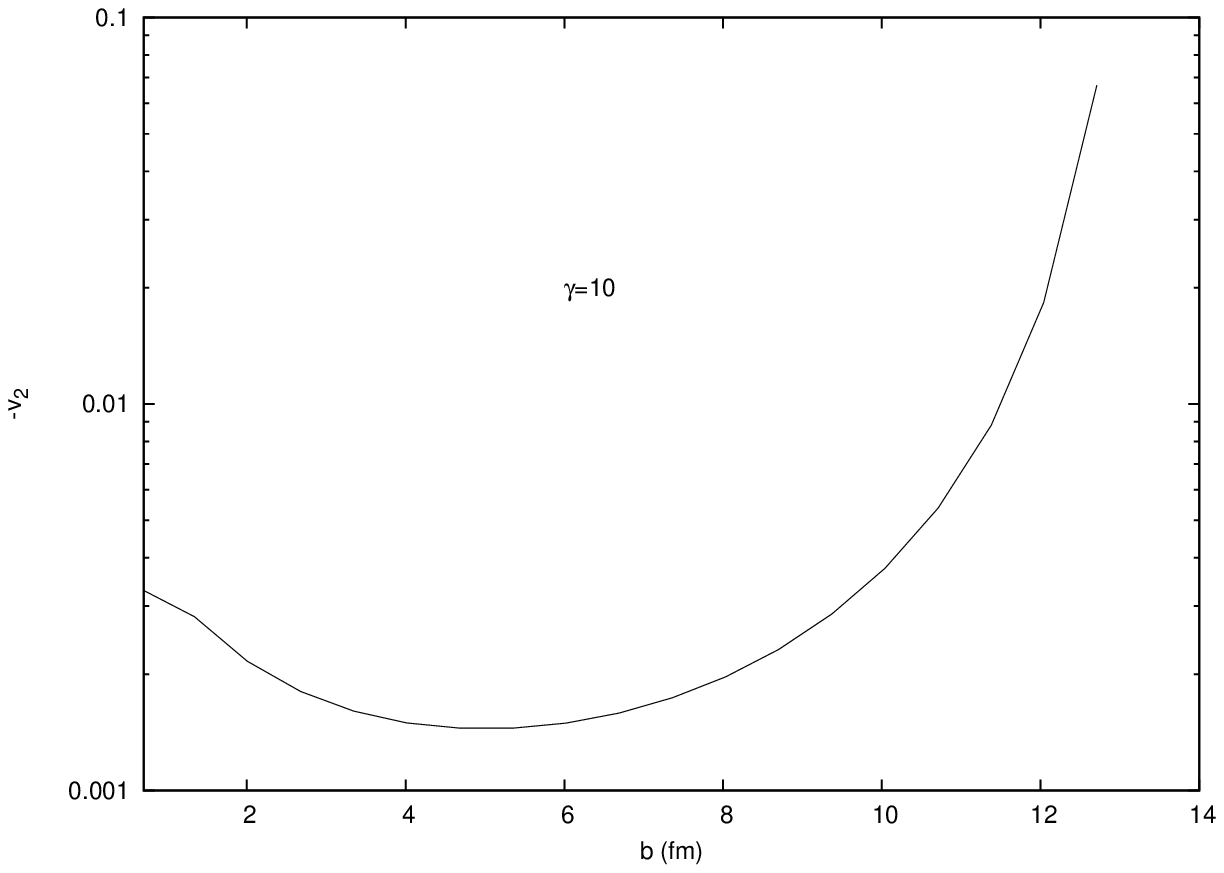,width=16 cm}
\caption{Elliptic flow coefficient $|v_2|$ with $\gamma=10$ as a function of $b$
for Au-Au collisions at $y=10$. }
\label{fig4}
\end{figure}

In (\ref {fig4}) we show the elliptic flow for  $\gamma=10$, when the resulting $v_2$ is negative. Its values turn out to be practically independent of $p$.
The magnitude  of $|v_2|$ and its behaviour with $b$ are very similar to what we have
obtained for values of $\gamma$ smaller than unity. Again   $|v_2|$
reaches value of the order of several percents only at very peripheral 
collisions.

\section{String fusion as a source of elliptic flow}
\subsection{The model}
As we have seen, propagation of strings in the transverse plane gives a certain
contribution to the elliptic flow, but it is too small compared to the experimental data and can only be noticeable at extremely peripheral collisions. 
So we pass to another source of elliptic flow related to the process of string fusion and mentioned in the Introduction. In fact this process is of the same nature as implied in the hydrodynamical approach, a sort of collective interaction. The difference is rather
quantitative, since fusion of strings brings the nuclear matter in the overlap
to the liquid state gradually, depending on the so-called percolation parameter $\rho$, which is  the string density in the transverse space. Only at values
$\rho>1.1\div 1.2$ drops of nuclear matter (fused strings) on the nuclear scale begin to form and at very high values of $\rho$ they tend to occupy all the overlap space.

The elliptic flow may have its origin in the fact that fused strings are
not symmetric in the transverse plane and therefore may emit particle with
different probability in different azimuthal directions. As a source of this asymmetry one may consider quenching of emitted particles as they pass through the
fused string area. In this case if a particle is emitted from the forward
surface of the fused string it will have a greater momentum than the one emitted from the backward surface, the latter having to pass through the string
loosing its energy. Generally, if the emitted particle has to propagate inside the string a path of length $l$ we may assume the probability to see the particle
of momentum $p$ outside the string to be proportional to
\beq 
P(p,l)=e^{-\frac{\xi p^2}{1-\sigma l}},
\label{probl}
\eeq
where $\sigma$ characterizes the loss of the transverse energy per unit length.
If the fused string is not symmetric, then the length $l$ will depend on the direction of emission, that is on the angle between the $x$ -axis related to the string and momentum ${\bf p}$. 

By itself this anisotropy of particle emission from fused strings cannot give any elliptic flow. If the overlap is azimuthally symmetric (impact parameter $b=0$) then obviously strings will occupy arbitrary directions respective to 
${\bf b}$
and the resulting distribution in momenta will be fully isotropic. However if the overlap is not symmetric in the azimuthal angle, then fused strings of different directions will form with different probabilities. This will give rise to a non-zero elliptic flow. A clear limiting case is when $\rho>>1$ and  all strings fuse into one which occupies all the overlap area and so is unsymmetric if the
latter is. 

Below we shall try to develop a more quantitative way to calculate the elliptic flow from string fusion. As a typical form of the fused string we shall assume 
a symmetric almond simlar to the shape of  the nuclear overlap for collisions
 of two identical nuclei with radius $R$ at distance $b$ between their centers.
It  is described by the equation
\beq \Big(x\pm\frac{b}{2}\Big) ^2+y^2=R^2
\label{eq2}
\eeq
where $x$-axis is directed along its minor axis.
The two  axes of the almond itself  $a$ and 
$d$ with $a>d$ are 
\beq
a=\sqrt{4R^2-b^2},\ \ d=2R-b,\ \ b\leq 2R.
\eeq 

\begin{figure}
\epsfig{file=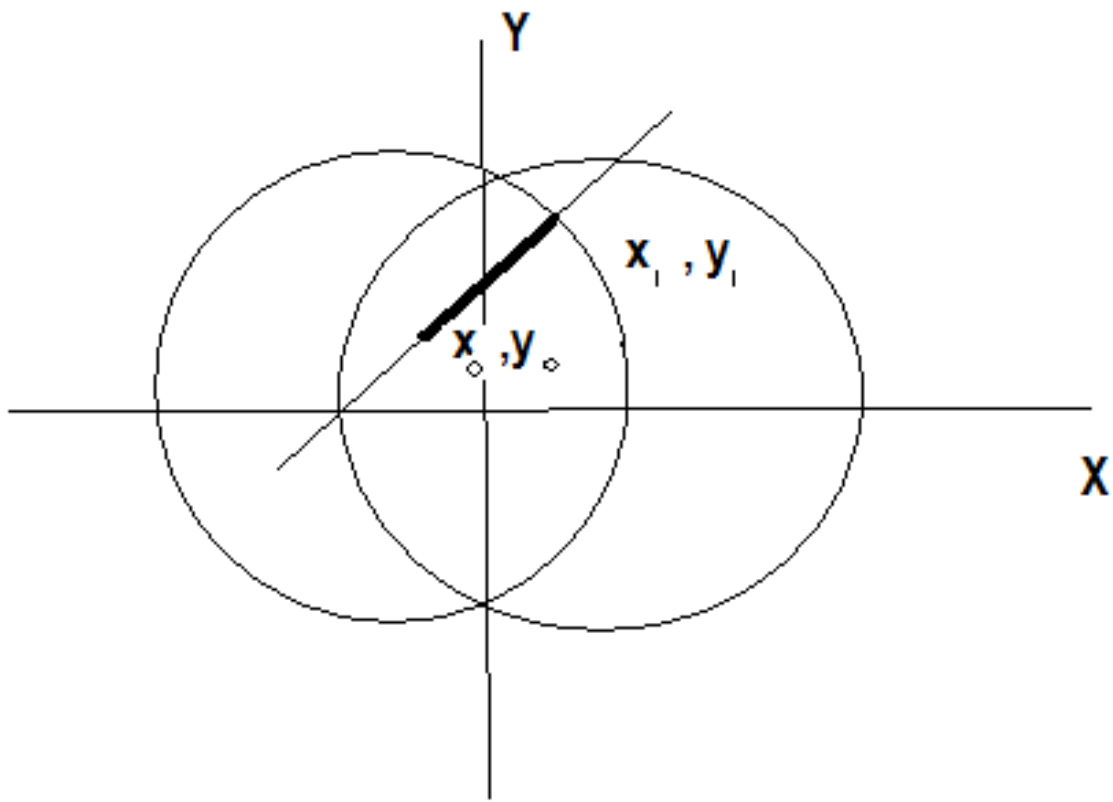,width=16 cm}
\caption{The almond of the nuclear overlap and the path of a particle emitted 
from point $(x_0,y_0)$ at a given angle to the minor axis of the diamond}
\label{fig5}
\end{figure}

We shall be interested in the emission of particles at  angle $\chi$ to the 
minor axis of the almond ($x$-axis) in the forward direction. Because of  symmetry we can assume $0\leq\chi\leq\pi/2$. Particles emiited at point $(x_0,y_0)$ in this direction move along the line
\beq y=\alpha x +c,\ \ c=y_0-\alpha x_0,\ \ \alpha=  \tan\chi
\label{eq1}
\eeq
(see Fig. \ref{fig5}).

 Points $x_1$ and $x_2$ on the surface of the string from which the particle is emitted in the forward and backward directions respectively
are obtained as solutions of the system of equations (\ref{eq1}) and
 (\ref{eq2}) with signs '$+$' and '$-$ respectively.
They may lie either on different sides of the almond or on the same side
depending on the values of $c$ and $\alpha$ 
Trivial manipulations
give 
\beq
x_1=\frac{1}{\lambda^2}\Big(-(+)\frac{b}{2}-\alpha c+\sqrt{Q_{1(2)}}\Big),\ \
 c<(>)\frac{a}{2} 
\label{x1}
\eeq
and
\beq
x_2=\frac{1}{\lambda^2}\Big(+(-)\frac{b}{2}-\alpha c-\sqrt{Q_{2(1)}}\Big),\ \ 
c>(<)-\frac{a}{2}, 
\label{x2}
\eeq
where 
\beq
\lambda^2=1+\alpha^2,\ \ Q_1=\lambda^2R^2-(c-\alpha b/2)^2,\ \ 
 Q_2=\lambda^2R^2-(c+\alpha b/2)^2.
\eeq

The length travelled by the emitted particle in the forward direction is given
by
\beq
l=\sqrt{x_1-x_0)^2+(y_1-y_0)²}=\lambda (x_1-x_0)=
\frac{\lambda}{\alpha}(y_1-y_0).
\label{ll}
\eeq

The emission probability will be given by (\ref{probl}). It will depend on the initial emission point $(x_0,y_0)$ and angle $\chi$.
To find the total emission probability at given $p$ and $\chi$ one has to 
integrate over all
$(x_0,y_0)$ in the almond area. Up to a constant factor
\beq
P(p,\chi)=\int dx_0dy_0e^{-\frac{\xi p^2}{1-\sigma l(\chi,x_0,y_0)}}.
\eeq
The total emission probability at given $\chi$ will be given after integration over all $p$. Again up to a constant factor
\beq
P(\chi)=\int dx_0dy_0\Big(1-\sigma l(\chi,x_0,y_0)\Big).
\eeq

Note that the expression for $l$ considerably simplifies for the extreme angles $\chi=0$ and $\chi=\pi/2$.
If $\chi=0$ then $\lambda=1$ and
\beq
x_1=-x_2=\sqrt{R^2-y_0^2}-\frac{b}{2}.
\eeq
If $\chi=\pi/2$ then $\lambda/\alpha=1$ and
\beq
y_1=-y_2=\sqrt{R^2-(|x_0|+b/2)^2}.
\eeq 
In these cases one can obtain $P(\chi)$ in an analytical 
form:
\beq
P(\chi=0)=\Omega-\sigma\Big[\frac{2}{3}a (2R^2+b^2)+
bR^2\Big(2\zeta+\sin(2\zeta)\Big)\Big]
\label{p0}
\eeq
and
\beq
P(\chi=\pi/2)=\Omega-\frac{1}{3}\sigma[8R^3-6bR^2+b^3/2\Big],
\label{p1}
\eeq
where
\beq
\zeta=\arccos\frac{b}{2R}
\eeq
and $\Omega$ is the almond area
\beq
\Omega=R^2\Big(2\zeta-\sin(2\zeta)\Big).
\eeq 

As mentioned, this anisotropy of emission from an unsymmetric string has to be combined with the anisotropic distribution of strings in the anisotropic overlap.
Choose the system in which the $x$ axis is directed along the impact parameter.
Let the minor axis of the almond corresponding to the fused string be directed
at azimuthal angle $\theta$. Then particle emitted at angle $\chi$ respective 
to the minor axis of the almond will be emitted at azimuthal angle $\phi=
\chi+\theta$ respective to  the direction of the impact parameter. It is the $\phi$-dependence which is measured experimentally, so that in our formulas for the
emission probablility and length $l$ one has to put $\chi\to\chi-\theta$.
The fused string can be generally directed at any angle $\theta$ respective to the direction of ${\bf b}$ with the probability $D(\theta)$
So the final distribution in the transverse momentum of particles emitted from thsi fused string will be obtained as
\beq
\frac{d\sigma}{pdpd\phi}=C\int d\theta D(\theta)
P(p,\phi-\theta)
\eeq
and the distribution in $\phi$ as:
\beq
\frac{d\sigma}{d\phi}=C_1\int d\theta  D(\theta)
P(\phi-\theta).
\label{dphi}
\eeq
Obviously if the distribution of strings in $\theta$ is isotropic, then
integration over $\theta$ will eliminate any dependence on $\phi$.

If the overlap is fully symmetric ($b=0$) then obviously we cannot expect any
dependence of the string distribution in $\theta$. However if it is not then
such a dependence certainly arises.
 
To see this, take the percolation parameter $\rho$ to be  very large, so that
practically all strings in the overlap area fuse into a single cluster, which occupies the whole overlap volume. With an uasymmetric overlap the only possibility to form such a cluster is to direct its major and minor axes along the mjor and minor axes of the overlap. In other words the distribition of strings in 
$\theta$ acquires a $\delta$-function form
\beq
D(\theta)= \delta(\theta) 
\eeq
and then
\beq
\frac{d\sigma}{pdpd\phi}=CP\Big(p,\phi),
\label{dpphi0}
\eeq
\beq
\frac{d\sigma}{d\phi}=C_1
P\Big(\phi)
\label{dphi0}
\eeq
where the almond occupies the whole overlap area,so in $P$ one has to put
$R=R_A$, the radius of colliding nuclei, and $b$ the actual impact parameter.
The obtained asymmetry is quite similar to the one in the  well-known mechanims
in which the emitted
partcles are suppressed by final-state interactions as they pass through the overlap area \cite{capella}.

The string fusion mechanism allows to obtain something more: the dependence of
the suppression on the string density through the percolation parameter
taking not so high values. Then strings occupy only part of the overlap area
given by factor 
\beq
F(\rho)=1-e^{-\rho}.
\eeq
This gives them a possibility to form different clusters and have different directions in the overlap area.
If $\rho$ and so the string length $l$ become small all directions become
equally possible, so that the distribution in $\theta$ becomes flat and with
that the distribition in $\phi$.

Unfortunately to study this $\rho$-dependence accurately
one has to recur to a very complicated numerical  Monte-Carlo methods 
to be able to see forms and directions of the string clusters in detail. 

To avoid this we may use
a crude simplified approach. We may assume that strings (fused and simple)
on the average are distributed homogeneously in the overlap area.
Again on the average, an emitted particle has to pass through the string matter
length $l$ which is smaller than the corresponding length at $\rho>>1$ by factor\beq
\kappa_1(\rho)=\sqrt{F(\rho)}.
\eeq  
The average string tension will be  greater than for the
single string by factor $\kappa(\rho)^{-1}$
where
\beq
\kappa(\rho)=\sqrt{\frac{1-e^{-\rho}}{\rho}}=\frac{\kappa_1}{\sqrt{\rho}}.
\eeq
The distribution in the transverse momentum will then be given by the same formulas (\ref{dpphi0}) and (\ref{dphi0}) with the appropriate rescaling of $l$ and
 $\xi$
\beq
\frac{d\sigma}{pdpd\phi}=C
P\Big(p,\phi, \xi\to \xi \kappa, l\to \kappa_1 l)=
C\int dx_0dy_0e^{-\frac{\kappa(\rho) \xi p^2}{1-\sigma \kappa_1(\rho)l(\phi,x_0,y_0)}}
\label{dpphi1}
\eeq
\beq
\frac{d\sigma}{d\phi}=C_1
P\Big(\phi, a\to a\kappa, l\to \kappa_1 l)
=\frac{C_1}{\kappa(\rho)}\int dx_0dy_0\Big(1-\kappa_1(\rho)
\sigma l(\phi,x_0,y_0)\Big)
\label{dphi1}
\eeq
This approach will be referred to as the 'averaged model'.

Note that  distribution (\ref{dpphi1}) formally may be interpreted
as following from the dependence of the percolation parameter $\rho$ on the
point inside the overlap.
Indeed one can rewrite  (\ref{dpphi1}) in the form
\beq
\frac{d\sigma}{pdpd\phi}=
C\int dx_0dy_0e^{-\kappa \rho_{eff}(\phi,x_0,y_0) \xi p^2},
\eeq
where the point- and direction-dependent effective percolation parameter
is determined by the obvious relation
\beq
\sqrt{
\frac{\rho_{eff}(\phi,x_0,y_0)}{1-e^{-\rho_{eff}(\phi,x_0,y_0)}}
}=
\frac{\sqrt{\rho}}{\sqrt{1-e^{-\rho}}}-\sigma\sqrt{\rho}l (\phi,x_0,y_0).
\eeq
In this way this picture coincides with the one proposed in \cite{bautista1}
and may be considered as its justification.

Calculations become especially simple if we neglect higher harmonics in the
expansion in $\cos(n\phi)$ and restrict to $n=2$.  Then we can obtain the
elliptic flow coefficient by comparing emission at angles $\phi=0$ and
$\phi=\pi/2$. In this case for the distribution integrated over $p$ we get an explicit expression
\[
2v_2=\frac{P(0)-P(\pi/2)}{P(0+P(\pi/2)}\]\beq=
\frac{1}{2}\frac{\epsilon(\rho)
\Big[8-6b+b^3/2-2a (2+b^2)-
3b\Big(2\zeta+\sin(2\zeta)\Big)
\Big]}
{2-\epsilon(\rho)
\Big[8-6b+b^3/2+2a (2+b^2)+
3b\Big(2\zeta+\sin(2\zeta)\Big)\Big]},
\eeq
where 
\beq
\epsilon(\rho)=\frac{\kappa_1\sigma R_A}{3\Omega}
\eeq
and we have put $R_A\to 1$ everywhere else, so that
\beq
\Omega=2\zeta-\sin(2\zeta),\ \ \zeta=\arccos\frac{b}{2}
\eeq
and $a$ and  $b$ are to be measured in units $R_A$.

A more elaborate picture ('elaborate model'), leading to a non-trivial 
$\theta$-dependence of $D(\theta)$, can be based on the assumption
that the macroscopic string 
cluster has the same almond shape as the overlap itself 
but its dimension is reduced by factor $\kappa$. Then its direction
may vary and the distribution of strings in $\theta$ will not 
be so sharply peaked at $\theta=0$
Accordingly we assume the  length of the major axis of the string is
\beq
a=\kappa_1\sqrt{4R_A^2-b^2}\equiv\kappa_1 a_0
\label{sca}
\eeq
Its direction respective to the direction of ${\bf b}$~ will be given by 
$\tau=\theta+\pi/2$. Symmetry of the almond dictates that the distribution does not change if $\theta\to-\theta$ and $\theta\to\pi-\theta$.
So it is sufficient to consider the case when the major axis of the 
string has its angle $\tau$
contained in the interval between $0$ and $\pi/2$ 
which corresonds to $-\pi/2<\theta<0$. Values of $D(\theta))$ at other angles will be obtained by  symmetry.

The probability that the string of dimension $l$ and major axis forming angle
$\tau$ with the direction of ${\bf b}$ is proportional to the part of almond 
area $A$ in which such a string can lie. This area is given by the integral
over the almond area over points ${\bf x}$ and ${\bf x}'$ such that they lie on a line forming angle $\tau$ with the $x$-axis and with $|{\bf x-x'}|=l$.
From (\ref{ll}) we find
\beq
A=\int\int d^2{\bf x} d^2{\bf x}' \delta(x'-x-l/\lambda)
\delta(y'-y-\alpha l/\lambda)
\label{ar}
\eeq
where $\alpha=\tan\tau$ and $\lambda=\sqrt{1+\alpha^2}$. The average length $l$
of the string will be given by Eq. (\ref{sca}).
 The two-dimensional
integral (\ref{ar}) can be easily calculated numerically. 
Note, that as follows from these numerical calculations, with the growth
of $b$ and $\rho$, the distribuition proportional to $A$ rapidly
takes the form of the $\delta$-function, concentrated at $\tau=\pi/2$, that is at $\theta=0$, which returns us to the previously considered averaged approximation.
So using this more elaborate picture is only reasonable at relatively small
values of $b$ and $\rho$ 

\subsection{Numerical results}
Our model contains a single new parameter $\sigma$ which characterizes the loss of energy in passing through the string field. In fact by dimensional reasons 
$\sigma$ is proportional to the nuclear radius $R_A$
Taking $R_A=R_0A^{1/3}$ we find 
\beq
\sigma=\sigma_0A^{1/3}
\eeq
where $\sigma_0$ is a dimensionless and $A$ independent parameter to be
extracted from the experimental data.

\begin{figure}
\epsfig{file=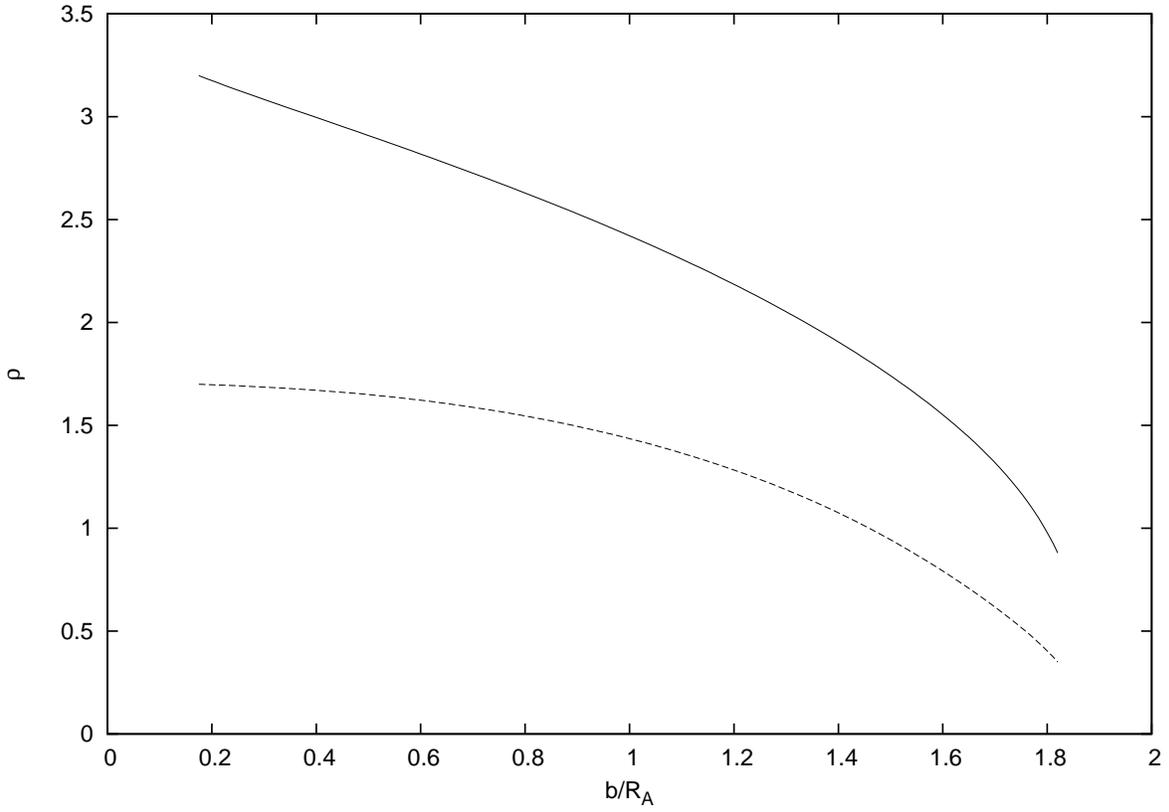,width=16 cm}
\caption{Percolation parameter as a function of impact parameter
for Au-Au collisions at 62.4 (lower curve) and 200 GeV
borrowed from \cite{tarn}}
\label{fig6}
\end{figure}

Values of the percolation parameter $\rho$ corresponding to given values of the impact parameter $b$ were taken from \cite{tarn} for Au-Au collisions at
62.4 and 200 GeV. They are shown in Fig. \ref{fig6}.
 
\begin{figure}
\epsfig{file=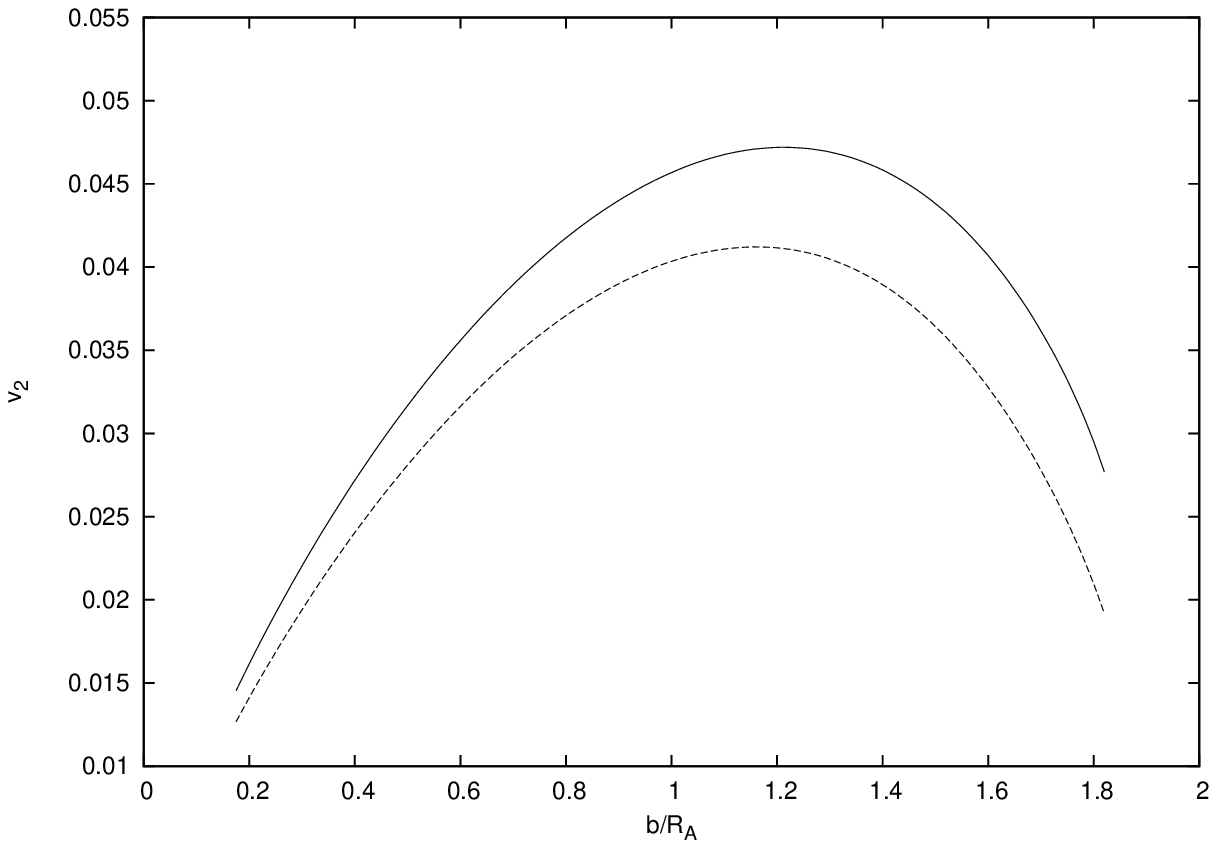,width=16 cm}
\caption{$v_2$ as a function of impact parameter
for Au-Au collisions at 62.4 (lower curve) and 200 GeV according
to Eq. (\ref{dphi1})}
\label{fig7}
\end{figure}

\begin{figure}
\epsfig{file=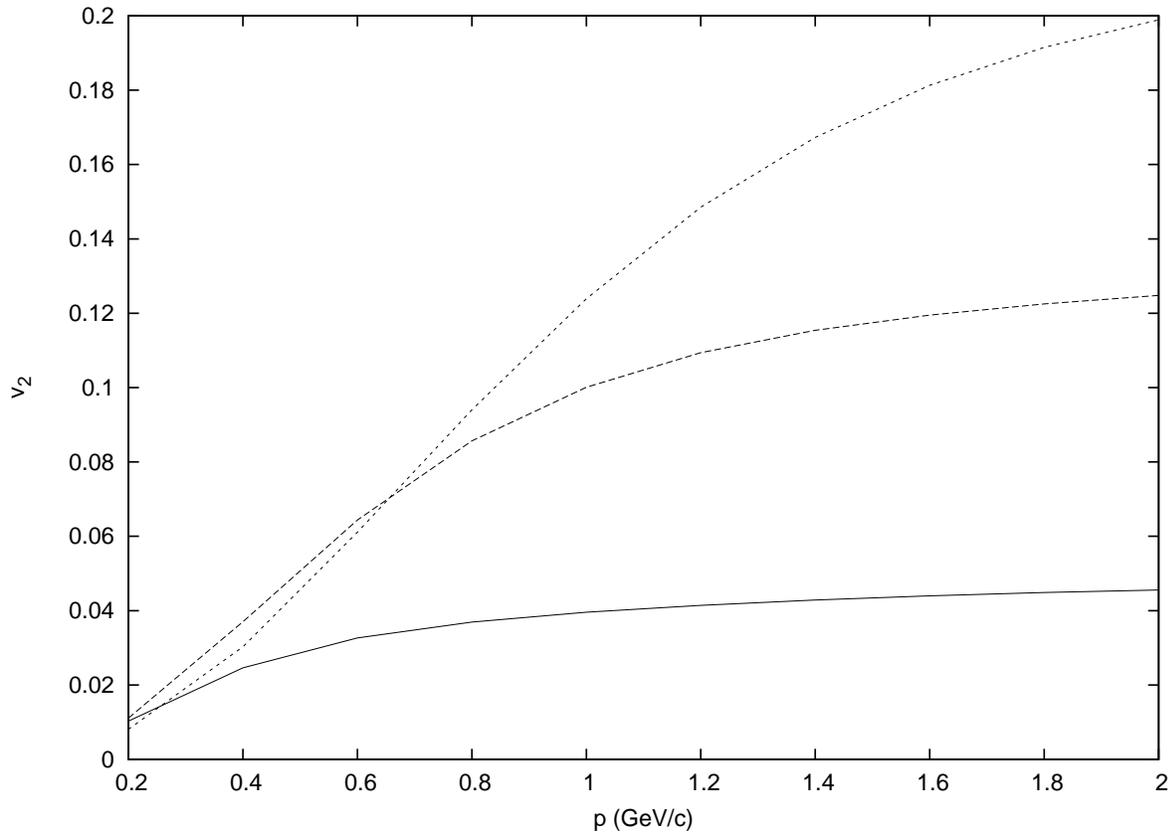,width=16 cm}
\caption{$v_2$ as a function of transverse momentum
for Au-Au collisions at  200 GeV according
to Eq. (\ref{dpphi1}) for peripheral (uppermost curve),
mid-central and central (lowest curve) collisions}
\label{fig8}
\end{figure}

Calculations of the elliptic flow coefficient $v_2$ from events integrated over
the transverse parameter as a function of impact parameter $b$ according
to the averaged formula (\ref{dphi1}) give the results shown in Fig \ref{fig7}
for Au-Au collisions at 62.4 and 200 GeV. 
Calulations according to the same
averaged picture (Eq. (\ref{dpphi1}) of the transverse momentum dependence
are presented in Fig \ref{fig8} for central, mid-central and peripheral
collisions. In both cases quenching parameter $\sigma_0$ was adjusted to the mid-central results for events integrated over transverse momenta. The adjusted value was $\sigma_0=0.09$
  
\begin{figure}
\epsfig{file=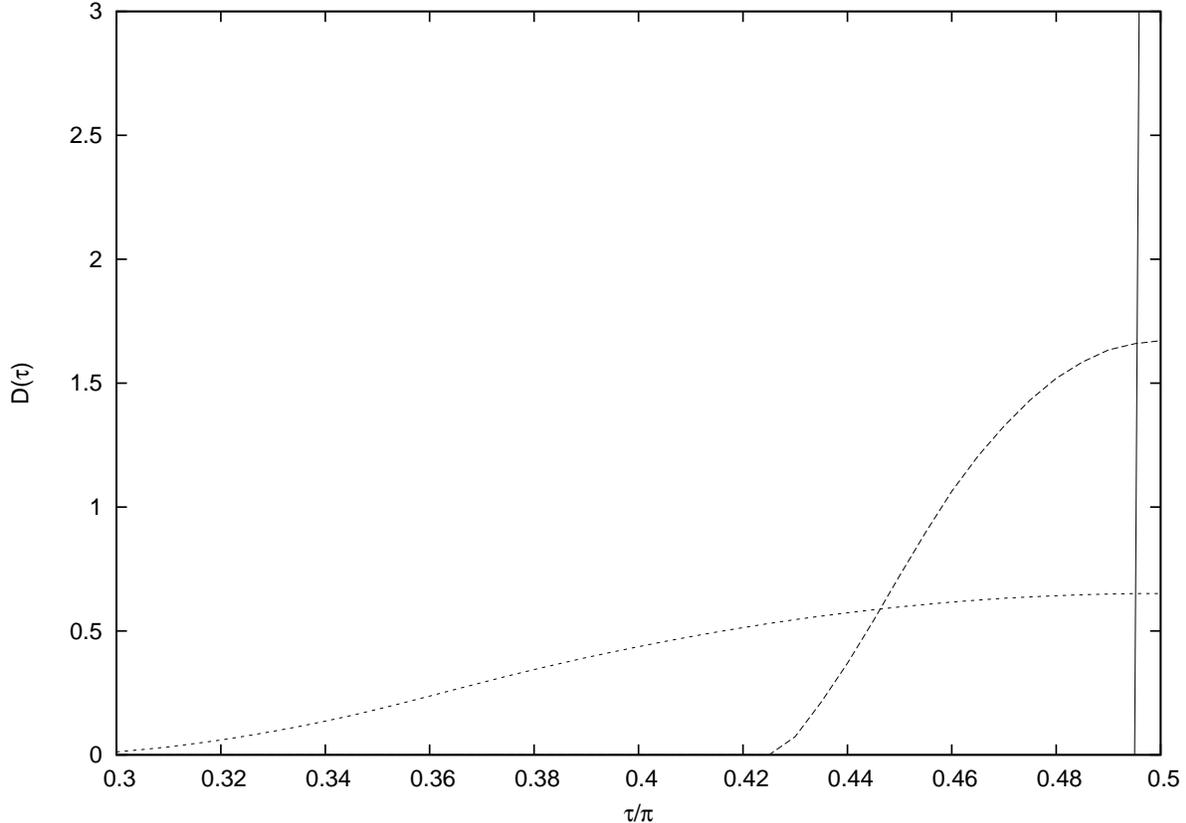,width=16 cm}
\caption{The distribution $D(\theta)$ as a function of $\tau=\pi/2-\theta$
for $b=R_A$ and $\rho=0.5$  (the leftmost curve), 1.0 (the middle curve) and
$\rho=2.$ (the rightmost curve, of which only the lowest part is shown) }
\label{fig9}
\end{figure}

Passing to our elaborate model, with a non-trivial distribution
in $\theta$, we, as mentioned, have found that with the growth of $\rho$
this distribution rapidly takes the form of a $\delta$ function. This is
illustrated in Fig. \ref{fig9} where we plot this distribution
as a function of angle $\tau=\pi/2-\theta$ for $b=R_A$ and $\rho=0.5,1$ and 2.
As one can see for $\rho=2$ the distribution is completely concentrated
at $\theta=0$
 
\begin{figure}
\epsfig{file=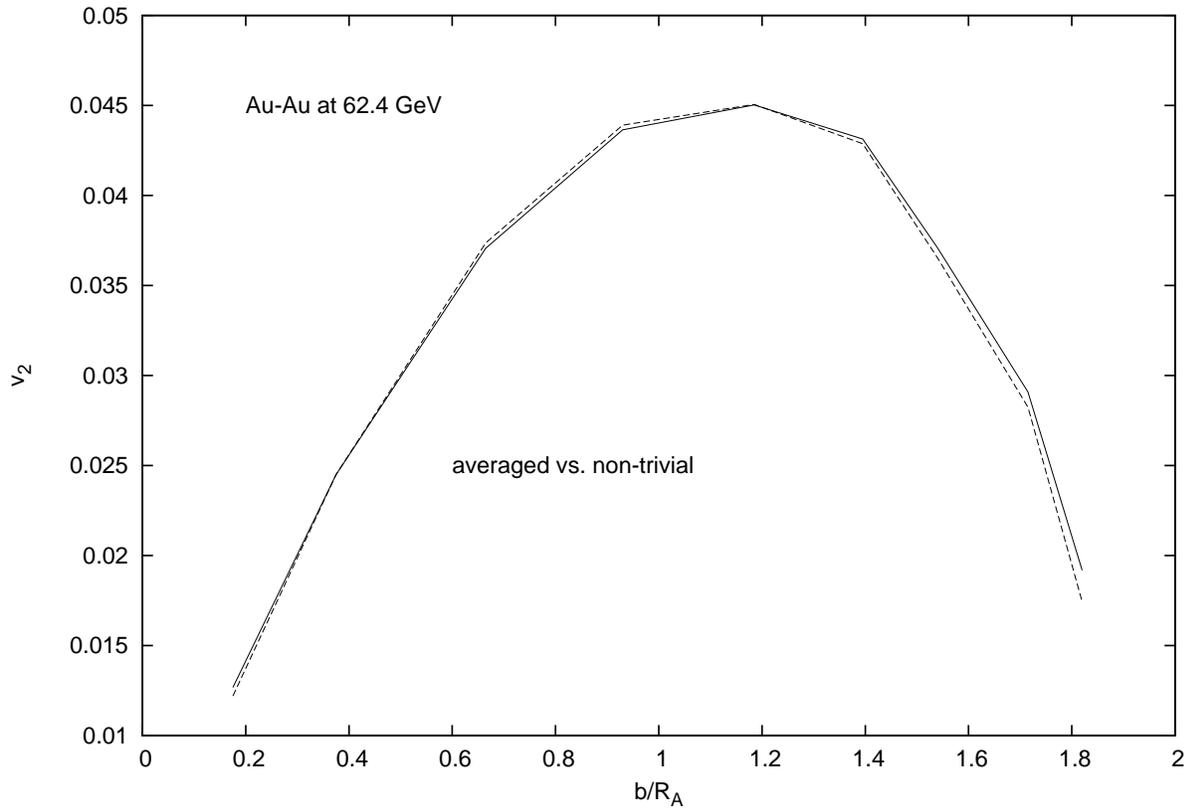,width=16 cm}
\caption{$v_2$ calculated in the averaged (solid curve )
 and elaborate models
for Au-Au collisions at 62.4 GeV as a function of $b$ with
$\sigma_0=0.09$}
\label{fig10}
\end{figure}

Turning to the data for $b$ and $\rho$ in Fig \ref{fig6} we find that for all
their values for Au-Au collisions at 200 GeV the distribution $D(\theta)$
is practically given by $\delta(\theta)$ so that our more elaborate model gives
nothing new as compared the previous crude one.
Calculations for Au-Au collisions at 62.4 GeV, for which the
distribution $D(\theta)$ is well different from a $\delta$-function
 give however results which are also practically identical with
 the averaged model. This is illustrated in Fig. \ref{fig10} in which we compare $v_2$ for events integrated over transverse momenta as a function of $b$ for Au-Au collisions at 62.4 GeV calculated according to Eqs. (\ref{dphi}) and
(\ref{ar}) Eq. (\ref{dphi1})
with $\sigma_0=0.09$. The difference is neglegible.
So the sophistication implied in the our elaborate model turns out to have  no practical value.

\begin{figure}
\epsfig{file=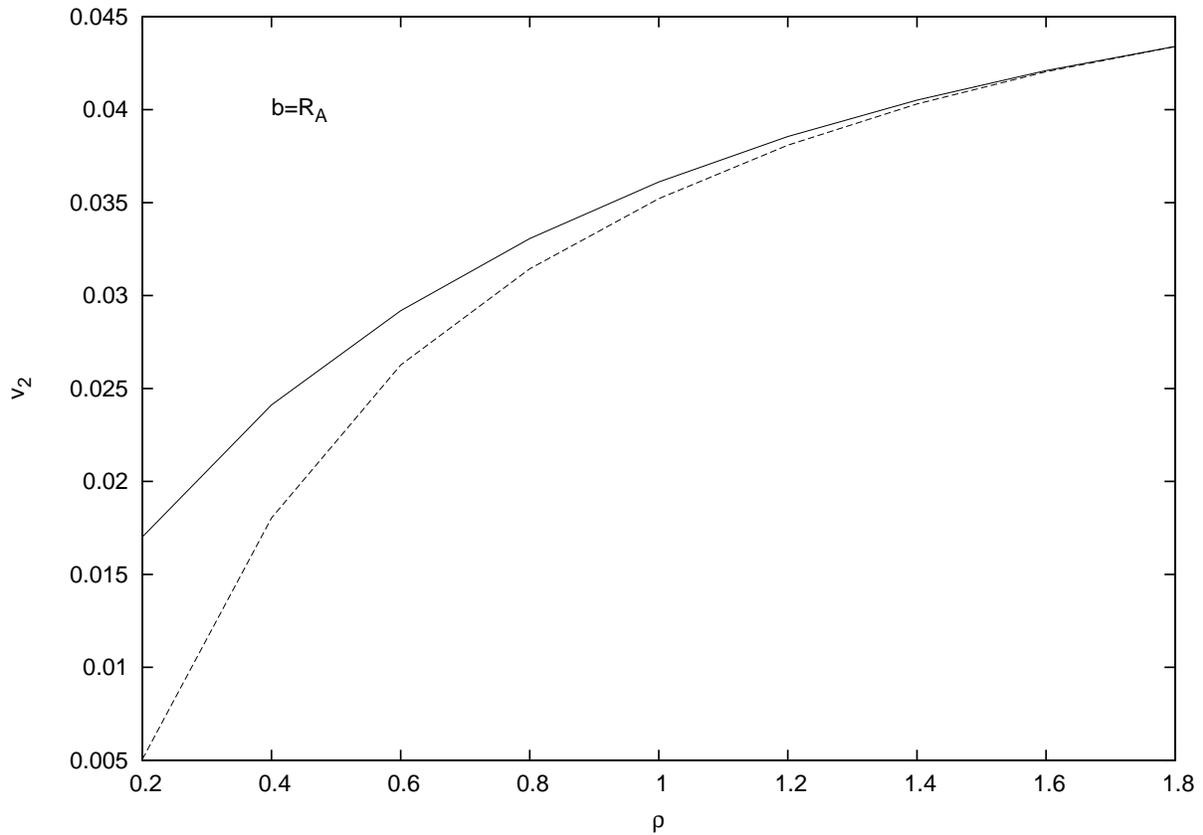,width=16 cm}
\caption{$v_2$ calculated in the averaged (solid curve)
 and elaborate models
for Au-Au collisions at fixed $b=R_A$ quenching parameter
$\sigma_0=0.09$ and $\rho$ changing from 0.2 to 1.8.}
\label{fig11}
\end{figure}

On the purely theoretical level one expects that the elaborate model places more emphasis on string fusion, so that it should give much less $\phi$ dependence
at low values of $\rho$ below the percolation threshold
$\rho\sim 1.1\div 1.2$. This is indeed so
as shows Fig. \ref{fig11} in which we plot $v_2$ for averaged and elaborate
models at fixed $b=R_A$ as a function of $\rho$ at compartively low values. One observes that at
values of $\rho$ below the percolation the non-trivial model gives $v_2$ substantially smaller than the averaged one.

\section{Conclusions}
We have demonstrated that the colour string model with fusion and percolation can succesfully describe the observed elliptic flow in high-energy heavy ion collisions. An important ingredient in this desription is  anisotropy of the string
emission spectra in the azimuthal direction. This may follow both from the
string propagation in the transverse palne due to a non-zero pomeron slope
and from quenching of the emitted partons in the strong colour field inside the string. We have found that the first source of anisotropy plays a minor role at accessible energies due to the fact that the distance travelled by the string in the transverse palne turns out to be small. The second source of anisotropy
however gives rise to anisotropy, which, upon adjusting the parameter of
quenching, allows to describe the data quite well both in their 
centrality dependence and their transverse momentum dependemce.

Our results have been obtained under some substantial approximations.
In the simplest case we used the over-all averaged picture both as to the
form of the string clusters as to their distribution in the nuclei overlap.
In the more elaborate case we fixed the geometric form of the leading cluster
and then found its distribution in the overlap. Both approximations have led
to practiclly the same results . 
Still carefull comparison demonstartes that the  leading cluster
approximation gives less flow at small values of the percolation paramter, which
indicates that string percolations is the most important source of the flow.
The results obtained here are similar to perevious calculations in 
\cite{bautista1,bautista2} in the same framework of percolation of strings under different approximations.

More accurate calculations of the flow in the string percolation model are only possible in the developed Monte-Carlo approach. They do not seem simple since one has to find an overall quenching for a given distribution of string clusters
in the overlap. We plan to conduct such calculations in future. 

\section{Acknowledgements}
This work is done under the projects FPA2008-01177 and Consolider of the 
Ministry of Science and Innovation of Spain and under the project of Xunta de Galicia. One of the autors (M.A.B) is indebted to the University of Santiago de Compostela for attention and financial support. He has also benefited from
grants RFFI 09-012-01327-a and RFFI-CERN 08-02-91004 of Russia which
partially supported this work.


\end{document}